\documentclass[journal]{IEEEtran}

\usepackage{amsmath}
\usepackage{cases}
\usepackage{amsfonts}
\usepackage{amssymb}
\usepackage{boxedminipage}
\usepackage{graphicx}
\usepackage[usenames]{color}
\usepackage{array,color}
\usepackage{colortbl}
\usepackage{lineno}
\usepackage{slashbox,pict2e}
\usepackage{bm}
\usepackage{algorithm}
\usepackage{algpseudocode}

\newtheorem{theorem}{\textbf{Theorem}}
\newtheorem{lemma}{\textbf{Lemma}}

\begin{document}

 \baselineskip=12pt
\title{ Power Allocation for Mixed Traffic Broadcast with Service Outage Constraint}
\author{Chuang~Zhang,~Pingyi~Fan\\  
State Key Laboratory on Microwave and Digital Communications\\
Department of Electronic Engineering, Tsinghua University, Beijing, P.R. China\\
E-mail:~zhangchuang11@mails.tsinghua.edu.cn,~fpy@tsinghua.edu.cn}
\maketitle

\begin{abstract}
To transmit a mixture of real-time and non-real-time traffic in a broadcast system, we impose a basic service rate $r_0$ for real-time traffic and use the excess rate beyond $r_0$ to transmit non-real-time traffic. Considering the time-varying nature of wireless channels, the basic service rate is guaranteed with a service outage constraint, where service outage occurs when the channel capacity is below the basic service rate. This approach is well suited for providing growing services like video, real-time TV, etc., in group transportation systems such as coach, high-speed train, and airplane. We show that the optimal power allocation policy depends only on the statistics of the minimum gain of all user channels, and it is a combination of water-filling and channel inversion. We provide the optimal power allocation policy, which guarantees that real-time traffic be delivered with quality of service (QoS) for every user. Moreover, we show that the required minimum average power to satisfy the service outage constraint increases linearly with the number of users.
\end{abstract}


\section{Introduction}

Mixed traffic, consisting of traffic with different delay requirements, permeates everywhere in today's communications. To differentiate the traffic, some previous works used the queueing delay, like \cite{dovrolis2002pdsddps} \cite{zhang2014pdsmsqsi}, however, deriving the queueing delay often requires the Markovian property of the queueing model, which might not be satisfied by practical traffic. To obviate such difficulty, in this paper, we broadly divide mixed traffic into two categories, real-time and non-real-time. Like in \cite{luo2003sobpra} \cite{cha2009sobprpfc}, we add a basic service rate $r_0$ for transmitting real-time traffic and use the excess rate beyond $r_0$ to transmit non-real-time traffic. If the channel capacity is smaller than $r_0$, service outage occurs. Quality of service (QoS) is guaranteed if the probability of service outage is smaller than a certain value $\epsilon$.


We use such a service outage-based approach for mixed traffic transmission in a broadcast system.  Specifically, we consider the scenario where the common transmitter needs to broadcast the same information to all users. This scenario is becoming more common with the development of group transportation systems such as coach, high-speed train, and airplane.\footnote{As in \cite{barbu2010}, the survey of European Space Agency's Advanced Research in Telecommunications Systems shows that ``broadband on trains" should include some real-time TV or personal multimedia services.} These services are more suitably provided by broadcasting over some reserved channel, since in most cases, other channels are allocated for control or Internet access. We will discuss power allocation in such a system to guarantee the service outage constraint of each user.

Our main contributions include: \emph{First,} we propose the optimal power allocation policy given service outage constraint in such a broadcast system, it turns out to be a combination of water-filling and channel inversion based only on the minimum gain of all user channels. \emph{Second,} we prove that the required minimum average power to guarantee QoS for each user scales linearly with the number of users. This result can serve as the upper bound of the power consumption and indicates that certain approaches like user cooperation or compromising fairness should be applied to avoid linear power consumption.

The rest of the paper is organized as follows. The system model and the main problem are introduced in Section \ref{sec_systmod}. In Sections \ref{sec_mainresult} and \ref{sec_solution}, we present the main theorem of the paper and the proof of this theorem respectively. In Section \ref{sec_smapnu}, we discuss the scaling of minimum average power with the number of users. Simulation results are given in Section \ref{sec_simulation}. Finally, conclusions are drawn in Section \ref{sec_conclusion}.

Notation: We use boldface letters to denote vectors, $\mathbb{E}_{\mathbf{h}\in\mathcal{H}}[\cdot]$ to denote expectation with random vector $\mathbf{h}$ in the state space $\mathcal{H}$, i.e., $\mathbb{E}_{\mathbf{h}\in\mathcal{H}}[\cdot]=\int_{\mathbf{h}\in\mathcal{H}}\cdot f_{\mathbf{h}}(\mathbf{h})\rm{d}\mathbf{h}$, where $f_{\mathbf{h}}(\mathbf{h})$ is the probability density function (PDF) of $\mathbf{h}$. If $\mathcal{H}$ is omitted, the expectation is over the entire state space of $\mathbf{h}$.

\section{System Model}\label{sec_systmod}

\begin{figure}
  \centering
  \includegraphics[width=0.6\columnwidth]{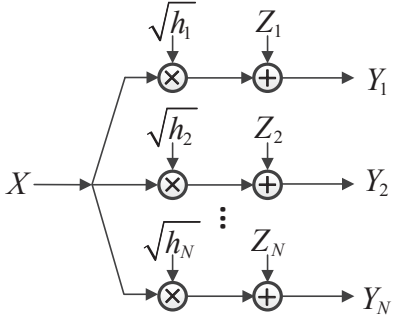}\\
  \caption{Broadcast channel model.}\label{fig_abssysmod}
\end{figure}

We discuss power allocation for a broadcast system in which a common transmitter provides the same service for $N$ users, as shown in Fig. \ref{fig_abssysmod}. The service is a mix of real-time and non-real-time traffic, we therefore impose a basic service rate $r_0$  for the real-time traffic. The excess rate beyond $r_0$ is utilized to transmit the non-real-time traffic.


Suppose all users in the system experience slow block fading, and the block length is long enough so that information-theoretic channel capacity can be applied. In a block, the channel of each user is
\begin{align}\label{eqn_sysmod}
Y_i=\sqrt{h_i}X+Z_i, \, i=1,2,\ldots,N,
\end{align}
where $\sqrt{h_i}$, $Z_i$ are the channel gain, additive white Gaussian noise of user $i$, respectively. Besides, the Gaussian noise of all users are assumed with the same variance $\sigma^2$.  If the transmitter has the average power constraint $P_{\text{av}}$, then the channel capacity in a block with channel gain $\sqrt{h_i}$ for user $i$ is denoted as $R(h_iP_{\text{av}})$, where $R(x)$ is
\begin{align}\label{eqn_chcap}
R(x)=\frac{1}{2}\log\Big(1+\frac{x}{\sigma^2}\Big),
\end{align}
and $\log$ is with base $2$ throughout this paper.

We define the instantaneous system capacity as the minimum instantaneous channel capacity of the $N$ users. If the instantaneous system capacity is below the basic service rate $r_0$, we call it a \emph{service outage}. Different from the concept of outage capacity, we allow variable-rate transmission by conducting either variable-rate channel coding or source coding. To guarantee QoS for all users, we constrain the probability of service outage to $\epsilon$, that is, the probability that the instantaneous system capacity is below $r_0$ is no larger than $\epsilon$. Furthermore, we assume that in each block, the common transmitter has perfect channel state information (CSI) of all users, then our main optimization problem is
\begin{align}
\max_{\gamma(\mathbf{h})}& \quad \mathbb{E}_{\mathbf{h}}[\min_i\{R(h_i\gamma(\mathbf{h}))\}]\label{eqn_basicpro}\\
\text{s.t.} & \quad \mathbb{E}_{\mathbf{h}}[\gamma(\mathbf{h})]\leq P_{\text{av}},\tag{\theequation a}\label{eqn_basicproa}\\
&\quad \gamma(\mathbf{h})\geq 0,\tag{\theequation b}\label{eqn_basicprob}\\
&\quad \text{Pr}\{\min_i\{R(h_i\gamma(\mathbf{h}))\}<r_0\}\leq \epsilon, \tag{\theequation c}\label{eqn_basicproc}
\end{align}
where $\gamma(\mathbf{h})$ is the power allocation policy, $\mathbf{h}$ is the channel state vector that $\mathbf{h}=(h_1,h_2,\ldots,h_N)$. Besides, we assume that $h_i, i=1,2,\ldots,N$ are continuous and independent with each other. (\ref{eqn_basicproa}) is the average power constraint, (\ref{eqn_basicprob}) requires power allocation to be nonnegative, and (\ref{eqn_basicproc}) is the service outage constraint.

By introducing $\check{h}=\min\{h_1, h_2,\ldots, h_N\}$, problem (\ref{eqn_basicpro}) can be reformulated as
\begin{align}
\max_{\gamma(\mathbf{h})}& \quad \mathbb{E}_{\mathbf{h}}[R(\check{h}\gamma(\mathbf{h}))]\label{eqn_basicproref}\\
\text{s.t.} & \quad \mathbb{E}_{\mathbf{h}}[\gamma(\mathbf{h})]\leq P_{\text{av}},\tag{\theequation a}\label{eqn_basicprorefa}\\
&\quad \gamma(\mathbf{h})\geq 0,\tag{\theequation b}\label{eqn_basicprorefb}\\
&\quad \text{Pr}\{\check{h}\gamma(\mathbf{h})<(2^{2r_0}-1)\sigma^2\}\leq \epsilon. \tag{\theequation c}\label{eqn_basicprorefc}
\end{align}

\section{Main Results} \label{sec_mainresult}

In this section, we present the main result on the optimal solution of problem (\ref{eqn_basicpro}), given as Theorem \ref{thm_mainthm}. The proof of this theorem is provided at the end of Section \ref{sec_solution}.

\begin{theorem} \label{thm_mainthm}
Define $f_{\check{h}}(\check{h})$ as the PDF of $\check{h}$, $\check{h}^{\epsilon}$ as the threshold that $\text{Pr}\{\check{h}\leq \check{h}^{\epsilon}\}=\epsilon$, $P_{\text{min}}$ as the minimum power that $P_{\text{min}}=\int_{\check{h}^{\epsilon}}^\infty \frac{(2^{2r_0}-1)\sigma^2}{\check{h}}f_{\check{h}}(\check{h})\rm{d}\check{h}$. Let $\gamma^*(\mathbf{h})$ be the optimal power allocation policy of problem (\ref{eqn_basicproref}) (or equivalently (\ref{eqn_basicpro})), then $\gamma^*(\mathbf{h})$ depends only on the statistics of the minimum of all channel gains. If $P_{\text{av}}< P_{\text{min}}$, $\gamma^*(\mathbf{h})$ does not exist. If $P_{\text{av}}\geq P_{\text{min}}$,
\begin{align}
\gamma^*(\mathbf{h})=\left\{\begin{array}{ll}
\min\{\lambda-\frac{1}{\check{h}},\frac{(2^{2r_0}-1)\sigma^2}{\check{h}}\}, & \check{h}\geq  \check{h}^{\epsilon},  \\{}
[\lambda-\frac{1}{\check{h}}]^+, & \text{otherwise,}
\end{array} \right.
\end{align}
where $\lambda$ is chosen that constraint (\ref{eqn_basicproa}) is satisfied.
\end{theorem}

The difficulty in proving Theorem 1 is that (\ref{eqn_basicprorefc}) is not a linear constraint of $\gamma(\mathbf{h})$, hence, concavity of $R(\check{h}\gamma(\mathbf{h}))$ with respect to (w.r.t.) $\gamma(\mathbf{h})$ is not enough. We need to show that all points of $\mathbf{h}$ with the same $\check{h}$ would either make $\check{h}\gamma(\mathbf{h})<(2^{2r_0}-1)\sigma^2$ or $\check{h}\gamma(\mathbf{h})\geq(2^{2r_0}-1)\sigma^2$ at the same time.

\section{Proof of Theorem \ref{thm_mainthm}} \label{sec_solution}

In this section, we start with the simple case where the number of users is $2$, and obtain the structure of the optimal solution, then extend the result to the case with $N$ users.

With 2 users, and by dividing the entire channel state space into two subspaces  $\mathcal{H}_1=\{\mathbf{h}|h_1\leq h_2\}$, $\mathcal{H}_2=\{\mathbf{h}|h_2< h_1\}$, we can rewrite problem (\ref{eqn_basicpro}) as
\begin{align}
\max_{\gamma(\mathbf{h})}& \quad \mathbb{E}_{\mathbf{h}\in\mathcal{H}_1}[R(h_1\gamma(\mathbf{h}))]+\mathbb{E}_{\mathbf{h}\in \mathcal{H}_2}[R(h_2\gamma(\mathbf{h}))] \label{eqn_basicpro2usersf2}\\
\text{s.t.} & \quad \mathbb{E}_{\mathbf{h}\in \mathcal{H}_1}[\gamma(\mathbf{h})]+\mathbb{E}_{\mathbf{h}\in\mathcal{H}_2}[\gamma(\mathbf{h})]\leq P_{\text{av}}, \tag{\theequation a}\label{eqn_basicpro2usersf2a}\\
&\quad \gamma(\mathbf{h})\geq 0,\tag{\theequation b}\label{eqn_basicpro2usersf2b}\\
&\quad \text{Pr}\{R(h_1\gamma(\mathbf{h}))< r_0, \mathbf{h}\in\mathcal{H}_1\}+\nonumber\\
&\quad \text{Pr}\{ R(h_2\gamma(\mathbf{h}))< r_0, \mathbf{h}\in\mathcal{H}_2\}\leq \epsilon. \tag{\theequation c}\label{eqn_basicpro2usersf2c}
\end{align}

Problem (\ref{eqn_basicpro2usersf2}) has the special structure that both the objective function and the constraints can be divided into two parts according to channel state space division. Therefore, we can divide this problem into two separate problems by introducing additional parameters. Let $\gamma_i(\mathbf{h})$ denote the power allocation policy in subspace $\mathcal{H}_i$, then the separation is given as (\ref{eqn_separation}).

\begin{align} \label{eqn_separation}
\max_{\{P_{1\text{av}}, P_{2\text{av}}, \epsilon_1,\epsilon_2\}}& \quad \mathbb{E}_{\mathbf{h}\in\mathcal{H}_1}[R(h_1\gamma_1(\mathbf{h}))]+\mathbb{E}_{\mathbf{h}\in\mathcal{H}_2}[R(h_2\gamma_2(\mathbf{h}))] \\
& \quad  \max_{\gamma_1(\mathbf{h})} \quad \mathbb{E}_{\mathbf{h}\in\mathcal{H}_1}[R(h_1\gamma_1(\mathbf{h}))] \label{eqn_sep1}\\
& \quad \quad \text{s.t.} \quad  \mathbb{E}_{\mathbf{h}\in\mathcal{H}_1}[\gamma_1(\mathbf{h})]\leq P_{1\text{av}}, \tag{\theequation a}\label{eqn_sep1a}\\
& \quad \quad \quad \quad \gamma_1(\mathbf{h})\geq 0,\tag{\theequation b}\label{eqn_sep1b}\\
&  \quad \quad \quad \quad \text{Pr}\{R(h_1\gamma_1(\mathbf{h}))< r_0, \mathbf{h}\in \mathcal{H}_1\}\leq \epsilon_1 \tag{\theequation c}\label{eqn_sep1c}\\
& \quad \max_{\gamma_2(\mathbf{h})} \quad \mathbb{E}_{\mathbf{h}\in\mathcal{H}_2}[R(h_2\gamma_2(\mathbf{h}))]\label{eqn_sep2}\\
& \quad\quad \text{s.t.}  \quad \mathbb{E}_{\mathbf{h}\in\mathcal{H}_2}[\gamma_2(\mathbf{h})]\leq P_{2\text{av}},\tag{\theequation a}\label{eqn_sep2a}\\
& \quad\quad \quad \quad \gamma_2(\mathbf{h})\geq 0,\tag{\theequation b}\label{eqn_sep2b}\\
& \quad\quad \quad \quad \text{Pr}\{R(h_2\gamma_2(\mathbf{h}))< r_0, \mathbf{h}\in \mathcal{H}_2\}\leq \epsilon_2 \tag{\theequation c}\label{eqn_sep2c}\\
& \quad P_{1\text{av}}+P_{2\text{av}}\leq P_{\text{av}}\\
& \quad \epsilon_1+\epsilon_2\leq \epsilon.
\end{align}

It is easy to check that the optimal solution of problem (\ref{eqn_separation}) is the same with that of problem (\ref{eqn_basicpro2usersf2}). (\ref{eqn_separation}) facilitates our analysis by dividing (\ref{eqn_basicpro2usersf2}) into two separate problems. By solving the two separate problems, we can find the optimal solution by adjusting the allocation of $P_{1\text{av}}, P_{2\text{av}}, \epsilon_1, \epsilon_2$.

Let's first take a look at the separate problem (\ref{eqn_sep1}). Without constraint (\ref{eqn_sep1c}), problem (\ref{eqn_sep1}) can be solved by using water-filling over $h_1$ in the channel state space $\mathcal{H}_1$. With constraint (\ref{eqn_sep1c}), we should further divide the channel state space $\mathcal{H}_1$ into two subspaces. Let $\mathcal{H}_1=\mathcal{H}_{1s}\cup\mathcal{H}_{1o}$, where $\mathcal{H}_{1s}=\{\mathbf{h}|R(h_1\gamma(\mathbf{h}))\geq r_0, \mathbf{h}\in \mathcal{H}_1 \}$, which we call the service subspace, and $\mathcal{H}_{1o}=\{ \mathbf{h}|R(h_1\gamma(\mathbf{h}))<r_0, \mathbf{h}\in\mathcal{H}_1\}$, which we call the outage subspace. Then power allocation in the channel state space $\mathcal{H}_1$ should make the probability of outage subspace no greater than $\epsilon_1$, that is $\text{Pr}\{\mathbf{h}\in\mathcal{H}_{1o}\}\leq\epsilon_1$. Let $h_1^{\epsilon_1}$ be the threshold for $h_1$ in the subspace $\mathcal{H}_1$ that $\int_{0}^{h_1^{\epsilon_1}}\int_{h_1}^{\infty} f_{\mathbf{h}}(\mathbf{h})\mathrm{d}h_2\mathrm{d}h_1=\epsilon_1$. Define $\mathcal{H}_1^{\epsilon_1}=\{\mathbf{h}|h_1\geq h_1^{\epsilon_1}, \mathbf{h}\in \mathcal{H}_1\}$, $\mathcal{\overline{H}}_1^{\epsilon_1}=\{\mathbf{h}|h_1<h_1^{\epsilon_1}, \mathbf{h}\in \mathcal{H}_1\}$, then to make the service outage constraint satisfied, we must have $\mathcal{H}_1^{\epsilon_1}\subset \mathcal{H}_{1s}$, for detailed proofs, see \cite{luo2003sobpra}. Then we can transform problem (\ref{eqn_sep1}) into the following form
\begin{align}
\max_{\gamma_1(\mathbf{h})}& \quad \mathbb{E}_{\mathbf{h}\in\mathcal{H}_1}[R(h_1\gamma_1(\mathbf{h}))] \label{eqn_separation1} \\
\text{s.t.} & \quad \mathbb{E}_{\mathbf{h}\in\mathcal{H}_1}[\gamma_1(\mathbf{h})]\leq P_{1\text{av}},\tag{\theequation a}\label{eqn_separation1a}\\
&\quad \gamma_1(\mathbf{h})\geq 0, \tag{\theequation b}\label{eqn_separation1b}\\
& \quad R(h_1\gamma_1(\mathbf{h}))\geq r_0, \mathbf{h}\in \mathcal{H}_{1}^{\epsilon_1}. \tag{\theequation c}\label{eqn_separation1c}
\end{align}

There is a minimum average power for problem (\ref{eqn_separation1}). According to constraint (\ref{eqn_separation1c}), $\gamma_1(\mathbf{h})\geq \frac{(2^{2r_0}-1)\sigma^2}{h_1}$, then $P_{\text{1av}}$ should satisfy
$$
P_{1\text{av}}\geq \int_{h_1^{\epsilon_1}}^{\infty}\int_{h_1}^{\infty}\tfrac{(2^{2r_0}-1)\sigma^2}{h_1}f_{\mathbf{h}}(\mathbf{h})\mathrm{d}h_2\mathrm{d}h_1.
$$


Let $P_{1\text{min}}=\int_{h_1^{\epsilon_1}}^{\infty}\int_{h_1}^{\infty}\frac{(2^{2r_0}-1)\sigma^2}{h_1}f_{\mathbf{h}}(\mathbf{h})\mathrm{d}h_2\mathrm{d}h_1$ be this minimum power, then if $P_{1\text{av}}<P_{1\text{min}}$, there is no power allocation policy which can meet the service outage constraint. If $P_{1\text{av}}\geq P_{1\text{min}}$,  we can solve this problem by using Lagrangian multiplier method.

Let the Lagrangian be
\begin{align} \nonumber
L\big(\gamma_1(\mathbf{h}),\tfrac{1}{\lambda_1 2\ln2}\big)=\mathbb{E}_{\mathbf{h}}\big[R(h_1\gamma_1(\mathbf{h}))-\tfrac{1}{\lambda_1 2\ln2}(\gamma_1(\mathbf{h})-P_{1\text{av}})\big].
\end{align}
By making the derivative of the Lagrangian $L$ over $\gamma_1(\mathbf{h})$ equal to $0$, we have
\begin{align} \nonumber
\frac{\partial L}{\partial \gamma_1(\mathbf{h})}=\mathbb{E}_{\mathbf{h}}\bigg[\frac{1}{2\ln2}\frac{h_1}{\sigma^2+h_1\gamma_1(\mathbf{h})}-\frac{1}{\lambda_1 2\ln2} \bigg]=0,
\end{align}
\begin{align} \nonumber
\Rightarrow \gamma_1(\mathbf{h})=\lambda_1 -\frac{\sigma^2}{h_1}.
\end{align}
On the other hand, $\gamma_1(\mathbf{h})$ must meet constraint (\ref{eqn_separation1c}) in the service subspace, therefore,
\begin{align} \label{eqn_sol2firstg}
\gamma_1(\mathbf{h})=\left\{\begin{array}{ll}
\max\Big\{\lambda_1-\frac{\sigma^2}{h_1}, \frac{(2^{2r_0}-1)\sigma^2}{h_1} \Big\}, & \mathbf{h}\in \mathcal{H}_{1}^{\epsilon_1},\\{}
[\lambda_1-\frac{\sigma^2}{h_1}]^+, & \mathbf{h}\in \mathcal{\overline{H}}_{1}^{\epsilon_1},
\end{array}
\right.
\end{align}
where $\lambda_1$ makes $\gamma_1(\mathbf{h})$ meet the power constraint (\ref{eqn_separation1a}).


Similarly, for the separate problem (\ref{eqn_sep2}), we have the power allocation
policy as given in the following.


If $P_{2\text{av}}\geq P_{2\text{min}}$, where $P_{2\text{min}}=\int_{h_2^{\epsilon_2}}^{\infty}\int_{h_2}^{\infty}\frac{(2^{2r_0}-1)\sigma^2}{h_2}f_{\mathbf{h}}(\mathbf{h})\mathrm{d}h_1\mathrm{d}h_2$,
\begin{align} \label{eqn_sol2secondg}
\gamma_2(\mathbf{h})=\left\{\begin{array}{ll}
\max\Big\{\lambda_2-\frac{\sigma^2}{h_2}, \frac{(2^{2r_0}-1)\sigma^2}{h_2} \Big\}, & \mathbf{h}\in\mathcal{H}_{2}^{\epsilon_2},\\{}
[\lambda_2-\frac{\sigma^2}{h_2}]^+, & \mathbf{h}\in\mathcal{\overline{H}}_{2}^{\epsilon_2},
\end{array}
\right.
\end{align}
where $\mathcal{H}_{2}^{\epsilon_2}=\{\mathbf{h}|h_2\geq h_2^{\epsilon_2}, \mathbf{h}\in \mathcal{H}_2\}$, $\mathcal{\overline{H}}_{2}^{\epsilon_2}=\{\mathbf{h}|h_2<h_2^{\epsilon_2}, \mathbf{h}\in \mathcal{H}_2\}$, $h_2^{\epsilon_2}$ is the threshold that $\int_{0}^{h_2^{\epsilon_2}}\int_{h_2}^{\infty} f_{\mathbf{h}}(\mathbf{h})\mathrm{d}h_1\mathrm{d}h_2=\epsilon_2$, and $\lambda_2$ makes $\gamma_2(\mathbf{h})$ meet the power constraint (\ref{eqn_sep2a}).

We've solved the separate problems (\ref{eqn_sep1}) (\ref{eqn_sep2}) respectively. To obtain the optimal solution of problem (\ref{eqn_separation}), it remains to determine the optimal value of $P_{1\text{av}}, P_{2\text{av}}, \epsilon_1, \epsilon_2$.

Regarding $(P_{1\text{av}}, P_{2\text{av}})$, we have the following theorem.

\begin{theorem} \label{thm_concavepwr2users}
Using the optimal power allocation policy  (\ref{eqn_sol2firstg})  (\ref{eqn_sol2secondg}), the objective function (\ref{eqn_separation}) is concave w.r.t. the average power vector $(P_{1\text{av}}, P_{2\text{av}})$.
\end{theorem}

The proof of Theorem \ref{thm_concavepwr2users} relies on the following lemmas.

\begin{lemma} \label{lem_concavecom}
Let $g$ be a function that $g(x):\mathbf{R^n} \mapsto \mathbf{R^k}$, $f$ be a function that $f(y):\mathbf{R^k}\mapsto \mathbf{R}$, then the composition of $g$ and $f$, $\phi(x)=f\circ g(x)=f(g_1(x),g_2(x),\cdots,g_k(x))$ is concave w.r.t. $x$ if $f(y)$ is concave w.r.t. $y$, $f(y)$ is nondecreasing in each argument, and $g_i(x)$ is concave w.r.t. $x$. This also applies for the case where $k\rightarrow \infty$.
\end{lemma}

See the Appendix for the proof.

\begin{lemma} \label{lem_concavepwruser1}
The objective function of problem (\ref{eqn_sep1}) $\mathbb{E}_{\mathbf{h}\in \mathcal{H}_1}[R(h_1\gamma_1(\mathbf{h}))]$ is a concave function of $P_{1\text{av}}$ if power is allocated using the policy  (\ref{eqn_sol2firstg}).
\end{lemma}

See the Appendix for the proof.

\begin{lemma} \label{lem_addconcave}
If two functions $f_1, f_2: \mathbf{R^n}\mapsto \mathbf{R}$ are concave, then the function $f: \mathbf{R^{2n}} \mapsto \mathbf{R}, f(x_1, x_2)=f_1(x_1)+f_2(x_2)$ is concave w.r.t. $(x_1,x_2)$. This follows from \cite{boydcvxopt} Section 3.2.1 and Section 3.2.2.
\end{lemma}

Based on Lemma \ref{lem_concavecom} - \ref{lem_addconcave}, we can prove Theorem \ref{thm_concavepwr2users}, see the Appendix for the proof.

Since the objective function is concave w.r.t. $(P_{1\text{av}},P_{2\text{av}})$, and $P_{1\text{av}}+P_{2\text{av}}=P_{\text{av}}$, we can utilize some convex optimization algorithms to search the optimal power vector for a certain $(\epsilon_1, \epsilon_2)$. However, it remains to determine the value of $(\epsilon_1, \epsilon_2)$.

Regarding $(\epsilon_1, \epsilon_2)$, we have the following theorem.

\begin{theorem} \label{thm_equalthr2users}
The optimal choice of $(\epsilon_1, \epsilon_2)$ to problem (\ref{eqn_separation}) must make $h_1^{\epsilon_1}=h_2^{\epsilon_2}$.
\end{theorem}

See the Appendix for the proof.

With Theorems \ref{thm_concavepwr2users} and \ref{thm_equalthr2users}, we can
solve problem (\ref{eqn_separation}) (equivalently problem (\ref{eqn_basicpro2usersf2})) by first determining the optimal division $\epsilon_1^*, \epsilon_2^*$ and then search for the optimal division of power. However, this method is difficult for more than 2 users since we cannot obtain the derivatives of the objective function. Then we try to reduce the $N$-dimensional problem into $1$-dimensional problem based on the above theorems.





Using similar discussions for the two-user case, we can also divide the channel state space $\mathcal{H}=\{\mathbf{h}\}$ into $N$ subspaces $\mathcal{H}=\bigcup_{i=1,2,\ldots,N}\mathcal{H}_i$, where $\mathcal{H}_i=\{\mathbf{h}|\min\{h_1,h_2,\ldots,h_N\}=h_i\}$. Likewise, we can decompose $P_{\text{av}}$, $\epsilon$ into $P_{\text{av}}=\sum_{i=1}^N P_{i\text{av}}$, $\epsilon=\sum_{i=1}^N \epsilon_i$. Then the optimal solution in  subspace $\mathcal{H}_i$ with power constraint $P_{i\text{av}}$ and service outage constraint $\epsilon_i$ is the same as solutions  (\ref{eqn_sol2firstg}). With these solutions, similar results as Theorems \ref{thm_concavepwr}, \ref{thm_equalthr} in the case of $N$ users can be obtained.
\begin{theorem} \label{thm_concavepwr}
The objective function (\ref{eqn_basicpro}) is concave w.r.t. $(P_{1\text{av}},P_{2\text{av}},\ldots, P_{N\text{av}})$ if power allocation is based on  (\ref{eqn_sol2firstg}) in each subspace $\mathcal{H}_i, i=1, 2, \cdots, N$.
\end{theorem}

\begin{theorem} \label{thm_equalthr}
Using power allocation  (\ref{eqn_sol2firstg}) in each subspace $\mathcal{H}_i, i=1, 2, \cdots, N$, the optimal solution of  problem (\ref{eqn_basicpro}) must make $h_1^{\epsilon_1}=h_2^{\epsilon_2}=\cdots=h_N^{\epsilon_N}$.
\end{theorem}

The proofs of Theorems \ref{thm_concavepwr}, \ref{thm_equalthr} are similar as those of Theorems \ref{thm_concavepwr2users}, \ref{thm_equalthr2users}, respectively, and are omitted here.

With the above results, we can prove Theorem \ref{thm_mainthm}.

\begin{proof}
Let $f_{-h_i}(h_i)=(1-F_{h_1}(h_i))\cdots(1-F_{h_{i-1}}(h_i))f_{h_i}(h_i)(1-F_{h_{i+1}}(h_i))\cdots(1-F_{h_{N}}(h_i))$, then we can express the objective function (\ref{eqn_basicpro}) as
\small
\begin{align}
&\mathbb{E}_{\mathbf{h}}[\min_i\{R(h_i\gamma(\mathbf{h}))\}] \nonumber\\
=&\sum_{i=1}^N\int_{\mathbf{h}\in\mathcal{H}_i}R(h_i\gamma(\mathbf{h}))f_{\mathbf{h}}(\mathbf{h})\rm{d}\mathbf{h}\nonumber\\
=&\sum_{i=1}^N\int_0^\infty R(h_i\gamma(\mathbf{h})) f_{-h_i}(h_i) \rm{d}h_i \nonumber \\
=&\sum_{i=1}^N\int_0^\infty R(x\gamma((h_1,\ldots,h_{i-1},x,h_{i+1},\ldots,h_N)))f_{-h_i}(x)\rm{d}x \nonumber
\end{align}
\normalsize
Since $R(x\gamma(\mathbf{h}))$ is concave w.r.t. $\gamma(\mathbf{h})$, we have
\begin{align}
\sum_{i=1}^N &R(x\gamma((h_1,\ldots,h_{i-1},x,h_{i+1},\ldots,h_N)))f_{-h_i}(x)
\nonumber\\
\leq & R(x\tfrac{\sum_{i=1}^Nf_{-h_i}(x)\gamma((h_1,\ldots,h_{i-1},x,h_{i+1},\ldots,h_N))}{\sum_{i=1}^Nf_{-h_i}(x)})\sum_{i=1}^Nf_{-h_i}(x).\nonumber
\end{align}
Therefore,
\small
\begin{align}
&\sum_{i=1}^N\int_0^\infty R(x\gamma((h_1,\ldots,h_{i-1},x,h_{i+1},\ldots,h_N)))f_{-h_i}(x)\rm{d}x \nonumber\\
\leq&\int_0^\infty R(x\tfrac{\sum_{i=1}^Nf_{-h_i}(x)\gamma((h_1,\ldots,h_{i-1},x,h_{i+1},\ldots,h_N))}{\sum_{i=1}^Nf_{-h_i}(x)})\sum_{i=1}^N f_{-h_i}(x)\rm{d}x . \label{eqn_intineq}
\end{align}
\normalsize
From (\ref{eqn_intineq}), we can see that for the channel states $\mathbf{h}$ in different subspaces with the same $\check{h}$, allocating the same power at those states would achieve larger average capacity. Besides, from Theorem \ref{thm_equalthr}, those channel states would all be either in the service set or in the outage set. Hence, allocating the power $\tfrac{\sum_{i=1}^Nf_{-h_i}(x)\gamma((h_1,\ldots,h_{i-1},x,h_{i+1},\ldots,h_N))}{\sum_{i=1}^Nf_{-h_i}(x)}$  would still make the service outage constraint satisfied. Therefore, the optimal policy would allocate the same power to the channel states with the same $\check{h}$, regardless of which subspace they are in.

Then we can transform problem (\ref{eqn_basicpro}) into $1$-dimensional power allocation problem using $\check{h}$,
\begin{align}
\max_{\gamma(\check{h})}& \quad \mathbb{E}_{\check{h}}[R(\check{h}\gamma(\check{h}))]\label{eqn_simbasicpro}\\
\text{s.t.} & \quad \mathbb{E}_{\check{h}}[\gamma(\check{h})]\leq P_{\text{av}},\tag{\theequation a}\label{eqn_simbasicproa}\\
&\quad \gamma(\check{h})\geq 0,\tag{\theequation b}\label{eqn_simbasicprob}\\
&\quad \text{Pr}\{R(\check{h}\gamma(\check{h}))<r_0\}\leq \epsilon, \tag{\theequation c}\label{eqn_simbasicproc}
\end{align}
where $\check{h}=\min_{i}\{h_i\}$.

The PDF of $\check{h}$ can be derived as
\begin{align} \label{eqn_PDFmin}
f_{\check{h}}(\check{h})=&\sum_{i=1}^N\Big((1-F_{h_1}(\check{h}))\cdots(1-F_{h_{i-1}}(\check{h}))f_{h_i}(\check{h})\nonumber\\
&(1-F_{h_{i+1}}(\check{h}))\cdots(1-F_{h_{N}}(\check{h}))\Big).
\end{align}

Then the optimal solution of (\ref{eqn_simbasicpro}) (equivalently, (\ref{eqn_basicpro})) can be derived similarly as (\ref{eqn_separation1}), as given in Theorem \ref{thm_mainthm}.
\end{proof}

\section{Scaling of Minimum Average Power with Number of Users} \label{sec_smapnu}

In this section, we discuss how the required minimum average power would scale with the increase of number of users. We assume that the channels of all users are Rayleigh fading with the same average channel gain $\Omega_1=\Omega_2=\cdots=\Omega_N$. Then, using equation (\ref{eqn_PDFmin}), we can obtain the PDF of $\check{h}$, it is in fact Rayleigh fading with $\check{\Omega}=\frac{\Omega_1}{N}$ (Correspondingly, the PDF of $\check{h}$ is minus exponential). From $P\{\check{h}\leq \check{h}^{\epsilon}=\epsilon\}$, we obtain $\check{h}^{\epsilon}=-\check{\Omega}\ln(1-\epsilon)$, substitute it into the equation of minimum average power, we obtain
\begin{align}
\check{P}_{\text{min}}(N)&=\int_{\check{h}^{\epsilon}}^\infty \frac{\sigma^2(2^{2r_0}-1)}{\check{h}}f_{\check{h}}(\check{h})\rm{d}\check{h} \nonumber\\
&=\int_{-\check{\Omega}\ln(1-\epsilon)}^\infty \frac{\sigma^2(2^{2r_0}-1)}{\check{h}}\frac{1}{\check{\Omega}}e^{-\tfrac{\check{h}}{\check{\Omega}}}\rm{d}\check{h}\nonumber\\
&\stackrel{(a)}{=}N\int_{-\Omega_1\ln(1-\epsilon)}^\infty \frac{\sigma^2(2^{2r_0}-1)}{h}\frac{ 1}{\Omega_1}e^{-\tfrac{h}{\Omega_1}}\rm{d}h \nonumber \\
&=N\check{P}_{\text{min}}(1) \label{eqn_PmscaleN}
\end{align}
where $(a)$ uses the substitution $\check{h}=\tfrac{h}{N}$.

As can be seen from Eqn. (\ref{eqn_PmscaleN}), the minimum average power scales linearly with the number of users $N$. Note that in this case, the basic service rate $r_0$ and the service outage probability $\epsilon$ remains unchanged. With the minimum average power, power allocated in the outage region is zero, and every user would have the same service outage probability. This could be different when $P_{\text{av}}>P_{\text{min}}$, as power allocated in the outage region may make the rate of some users larger than $r_0$, hence, their service outage probability can be smaller.

The result of (\ref{eqn_PmscaleN}) can be used as the required power upper bound of this broadcast system when discussing user cooperation or fairness. It is the case where there is no cooperation among users at all or absolute fairness is achieved among all users. Any approach which allows cooperation (like relay, multiuser MIMO) or sacrifices a certain degree of fairness would have performance better than (\ref{eqn_PmscaleN}).

\section{Numerical Results} \label{sec_simulation}

In this section, we conduct simulations to justify the analysis. All users are subject to Rayleigh fading, and the average channel gain of user $i$ is $\Omega_i, i=1,2,\cdots,N$. Besides, Gaussian noise with normalized variance for each user is used.

\begin{figure}
  \centering
  \includegraphics[width=\columnwidth]{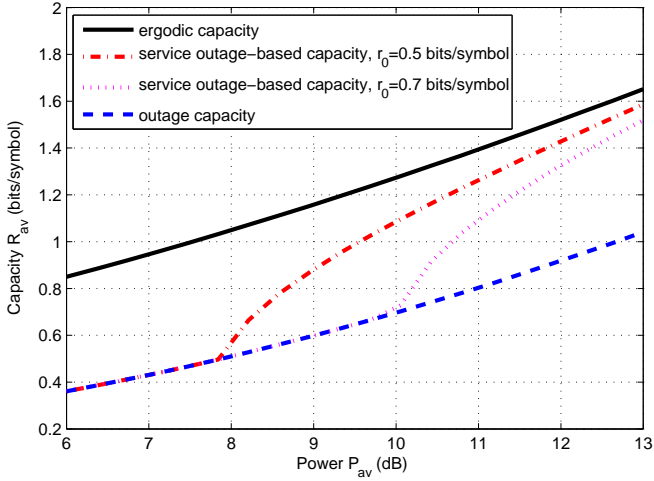}\\
  \caption{Comparison of service outage-based capacity with other capacities in a two-user Rayleigh fading channels, where $\Omega_1=1, \Omega_2=2$, $\epsilon=0.01$. }\label{fig_cap2users}
\end{figure}

Fig. \ref{fig_cap2users} is different capacities versus average power curves in a two-user broadcast system. One can observe that service outage-based capacity lies between ergodic capacity and outage capacity. As the average power increases, the service outage-based capacity increases from outage capacity to ergodic capacity. Since a basic service rate $r_0$ is guaranteed with probability $1-\epsilon$, the service outage-based capacity is lower than ergodic capacity. The increase of service outage-based capacity over outage capacity is due to variable-rate transmission. The difference between the service outage-based capacity and the basic service rate is the rate for sending non-real-time traffic. For instance, with average power $9$ dB and basic service rate $0.5$ bits/sysmbol, the rate for non-real-time traffic is approximately $0.37$ bits/symbol. Furthermore, when the basic service rate increases, the achievable service outage-based capacity decreases with the same average power. This is due to that more power are consumed for guaranteeing the higher service rate in poor channel states, and therefore, power is less efficiently used than that with a lower basic service rate.

\begin{figure}
  \centering
  \includegraphics[width=\columnwidth]{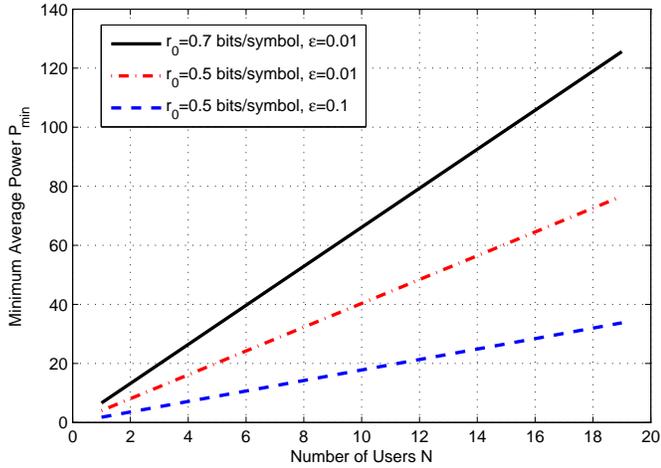}\\
  \caption{Scaling of minimum average power with the number of users, $\Omega_1=\Omega_2=\cdots=\Omega_N=1$. }\label{fig_minpwrscale}
\end{figure}

Fig. \ref{fig_minpwrscale} shows the scaling of minimum average power $P_{\text{min}}$ with the number of users $N$ in three different cases. As can be seen, $P_{\text{min}}$ increases linearly with $N$ in all three cases. This is actually the upper bound of the required minimum average power to meet the service outage constraint in such a broadcast system. If cooperation among users are allowed, for instance, using techniques like relay or multiuser MIMO, $P_{\text{min}}$ can decrease with the increase of $N$, which is due to diversity. Moreover, this system achieves absolute fairness among all users (from the perspective that each user is guaranteed the same service outage constraint), if fairness is compromised, $P_{\text{min}}$ can also decrease with $N$, which arises from multiuser diversity. Hence, the obtained scaling can be treated as an upper bound of minimum average power in a broadcast system with user cooperation or which compromises fairness with efficiency for mixed traffic transmission.

\section{Conclusion} \label{sec_conclusion}

In this work, we investigated transmitting mixed traffic in a broadcast system using the concept of service outage. We derived the optimal power allocation policy by reducing $N$-dimensional problem into $1$-dimensional problem. Furthermore, we illustrated that without cooperation among users, the required minimum average power to guarantee QoS for each user would increase linearly with the number of users. The linear scaling entails user cooperation or making a compromise between fairness and efficiency.

\appendix

\subsection{Proof of Lemma \ref{lem_concavecom}}

\begin{proof}
The proof follows from the definition of concavity. Let $x_1, x_2\in \text{dom}\ g(x)$, $0 \leq\alpha\leq 1$, and $ \overline{\alpha}=1-\alpha$, then we need to show that $\phi(\alpha x_1+\overline{\alpha} x_2)\geq \alpha\phi(x_1)+\overline{\alpha}\phi(x_2)$.

Since $g_i(x)$ is concave w.r.t. $x$, we have $g_i(\alpha x_1+\overline{\alpha} x_2)\geq \alpha g_i(x_1)+\overline{\alpha}g_i(x_2)$, $i=1,2,\ldots,k$. Therefore,
\begin{align*}
&\phi(\alpha x_1+\overline{\alpha} x_2)=f\circ g(\alpha x_1+\overline{\alpha} x_2)\\
\stackrel{(a)}{\geq} & f(\alpha g_1(x_1)+\overline{\alpha}g_1(x_2),\ldots,\alpha g_k(x_1)+\overline{\alpha}g_k(x_2)) \\
\stackrel{(b)}{\geq} & \alpha f(g(x_1))+\overline{\alpha} f(g(x_2))=\alpha\phi(x_1)+\overline{\alpha}\phi(x_2),
\end{align*}
where $(a)$ is due to that $f(y)$ is nondecreasing in each argument, $(b)$ is from the concavity of $f(y)$.

From the concavity proof, we can see that the composition of $f$ and $g$ preserves concavity when $k \rightarrow \infty$.
\end{proof}

\subsection{Proof of Lemma \ref{lem_concavepwruser1}}

\begin{proof}
We express $\mathbb{E}_{\mathbf{h}\in \mathcal{H}_1}[R(h_1\gamma_1(\mathbf{h}))]$ as
\begin{align}
&\mathbb{E}_{\mathbf{h}\in \mathcal{H}_1}[R(h_1\gamma_1(\mathbf{h}))] \nonumber\\
=&\int_{\mathbf{h}\in\mathcal{H}_1} \frac{1}{2}\log\bigg(1+\frac{h_1\gamma_1(\mathbf{h})}{\sigma^2}\bigg)f_{\mathbf{h}}(\mathbf{h})\mathrm{d} \mathbf{h}.
\end{align}
We see from the above formula that $\mathbb{E}_{\mathbf{h}\in \mathcal{H}_1}[R(h_1\gamma_1(\mathbf{h}))]$ is concave w.r.t. $\gamma_1(\mathbf{h})$ and nondecreasing at each point of $\gamma_1(\mathbf{h})$.

On the other hand, when $P_{1\text{av}}\geq P_{1\text{min}}$, we can reformulate the expression of (\ref{eqn_sol2firstg}) as
\begin{align} \label{eqn_pwrform}
\gamma_1(\mathbf{h})=P_{\text{CI}}(\mathbf{h})+P_{\text{res}}(\mathbf{h}),
\end{align}
where
\begin{align} \label{eqn_pwrci}
P_{\text{CI}}(\mathbf{h})=\left\{\begin{array}{ll}
\frac{(2^{2r_0}-1)\sigma^2}{h_1}, & \mathbf{h}\in\mathcal{H}_{1}^{\epsilon_1},\\
0, & \mathbf{h}\in \mathcal{\overline{H}}_{1}^{\epsilon_1},
\end{array}\right.
\end{align}
\begin{align} \label{eqn_pwrres}
P_{\text{res}}(\mathbf{h})=\left\{\begin{array}{ll}
[\lambda_1-\frac{2^{2r_0}\sigma^2}{h_1}]^+, & \mathbf{h}\in\mathcal{H}_{1}^{\epsilon_1},\\{}
[\lambda_1-\frac{\sigma^2}{h_1}]^+, & \mathbf{h}\in\mathcal{\overline{H}}_{1}^{\epsilon_1},
\end{array}\right.
\end{align}
In order to maximize the capacity, $\gamma_1(\mathbf{h})$ must meet the average power constraint with equality, that is
\begin{align}
\mathbb{E}_{\mathbf{h}\in \mathcal{H}_1}[\gamma_1(\mathbf{h})]
=\int_{\mathbf{h}\in\mathcal{H}_1}\gamma_1(\mathbf{h})f_{\mathbf{h}}(\mathbf{h})\mathrm{d}\mathbf{h}=P_{1\text{av}}.
\end{align}

Substitute the expressions of $P_{\text{CI}}(\mathbf{h})$ and $P_{\text{res}}(\mathbf{h})$ into the above equation, we have
\begin{align*}
\int_{\mathbf{h}\in\mathcal{H}_1}P_{\text{res}}(\mathbf{h})f_{\mathbf{h}}(\mathbf{h})\mathrm{d}\mathbf{h}=P_{1\text{av}}-P_{1\text{min}}.
\end{align*}
If we assume that the CSI of these two users are independent, then
\small
\begin{align*}
&\int_{\mathbf{h}\in\mathcal{H}_1}P_{\text{res}}(\mathbf{h})f_{\mathbf{h}}(\mathbf{h})\mathrm{d}\mathbf{h}\\
=&\int_{\min\{\frac{\sigma^2}{\lambda_1},h_1^{\epsilon_1}\}}^{h_1^{\epsilon_1}} \int_{h_1}^{\infty}\Big(\lambda_1-\tfrac{\sigma^2}{h_1}\Big)f_{h_1}(h_1)f_{h_2}(h_2)\mathrm{d}h_2\mathrm{d}h_1+\\
&\int_{\max\{\frac{2^{2r_0}\sigma^2}{\lambda_1},h_1^{\epsilon_1}\}}^{\infty} \int_{h_1}^{\infty}\Big(\lambda_1-\tfrac{2^{2r_0}\sigma^2}{h_1}\Big)f_{h_1}(h_1)f_{h_2}(h_2)\mathrm{d}h_2\mathrm{d}h_1
\end{align*}
\normalsize

For the general case, we assume that $\frac{\sigma^2}{\lambda_1}\leq h_1^{\epsilon_1}\leq \frac{2^{2r_0}\sigma^2}{\lambda_1}$, other situations can be discussed similarly. Let $F_{h_2}(x)=\int_{0}^{x}f_{h_2}(h_2)\mathrm{d}h_2$, then the above equation becomes
\small
\begin{align*}
&\int_{\mathbf{h}\in\mathcal{H}_1}P_{\text{res}}(\mathbf{h})f_{\mathbf{h}}(\mathbf{h})\mathrm{d}\mathbf{h}\\
=&\int_{\frac{\sigma^2}{\lambda_1}}^{h_1^{\epsilon_1}} \Big(\lambda_1-\tfrac{\sigma^2}{h_1}\Big)f_{h_1}(h_1)(1-F_{h_2}(h_1))\mathrm{d}h_1+\\
&\int_{\frac{2^{2r_0}\sigma^2}{\lambda_1}}^{\infty} \Big(\lambda_1-\tfrac{2^{2r_0}\sigma^2}{h_1}\Big)f_{h_1}(h_1)(1-F_{h_2}(h_1))\mathrm{d}h_1\\
=&P_{1\text{av}}-P_{1\text{min}}
\end{align*}
\normalsize
Let $\lambda_1$ be a function of $P_{1\text{av}}$, then after taking the derivative of both sides of the above equation, we have
\small
\begin{align*}
&\lambda_1^{'}\int_{\frac{\sigma^2}{\lambda_1}}^{h_1^{\epsilon_1}}f_{h_1}(h_1)(1-F_{h_2}(h_1))\mathrm{d}h_1+\\
&\lambda_1^{'}\int_{\frac{2^{2r_0}\sigma^2}{\lambda_1}}^{\infty}f_{h_1}(h_1)(1-F_{h_2}(h_1))\mathrm{d}h_1=0
\end{align*}
\normalsize
Taking the derivative of the above equation, we get
\small
\begin{align*}
&\lambda_1^{''}\bigg(\int_{\frac{\sigma^2}{\lambda_1}}^{h_1^{\epsilon_1}}f_{h_1}(h_1)(1-F_{h_2}(h_1))\mathrm{d}h_1 \\&+\int_{\frac{2^{2r_0}\sigma^2}{\lambda_1}}^{\infty}f_{h_1}(h_1)(1-F_{h_2}(h_1))\mathrm{d}h_1 \bigg)\\
=&-\sigma^2\frac{(\lambda_1^{'})^2}{\lambda_1^2}\bigg(f_{h_1}\Big(\frac{\sigma^2}{\lambda_1}\Big)\Big(1-F_{h_2}\Big(\frac{\sigma^2}{\lambda_1}\Big)\Big) +\\
&f_{h_1}\Big(\frac{2^{2r_0}\sigma^2}{\lambda_1}\Big)\Big(1-F_{h_2}\Big(\frac{2^{2r_0}\sigma^2}{\lambda_1}\Big)\Big)\bigg)
\end{align*}
\normalsize

From the above equation, we can see that $\lambda_1^{''}\leq 0$, therefore, $\lambda_1$ is concave w.r.t. $P_{1\text{av}}$.

Combining the expressions (\ref{eqn_pwrform}) (\ref{eqn_pwrci}) (\ref{eqn_pwrres}), we see that $\gamma_1(\mathbf{h})$ is concave w. r. t. $P_{1\text{av}}$ at each point of $\mathbf{h}$. According to Lemma 1, $\mathbb{E}_{\mathbf{h}\in \mathcal{H}_1}[R(h_1\gamma_1(\mathbf{h}))]$ is concave w.r.t. $P_{1\text{av}}$.
\end{proof}

\subsection{Proof of Theorem \ref{thm_concavepwr2users}}

\begin{proof}
According to Lemma \ref{lem_concavepwruser1}, $\mathbb{E}_{\mathbf{h}\in \mathcal{H}_1}[R(h_1\gamma_1(\mathbf{h}))]$ is concave w.r.t. $P_{1\text{av}}$, $\mathbb{E}_{\mathbf{h}\in \mathcal{H}_2}[R(h_2\gamma_2(\mathbf{h}))]$ is concave w.r.t. $P_{2\text{av}}$.  Then, based on Lemma \ref{lem_addconcave}, $\mathbb{E}_{\mathbf{h}\in \mathcal{H}_1}[R(h_1\gamma_1(\mathbf{h}))]+\mathbb{E}_{\mathbf{h}\in \mathcal{H}_2}[R(h_2\gamma_2(\mathbf{h}))]$ is  concave w.r.t. $(P_{1\text{av}},P_{2\text{av}})$.
\end{proof}

\subsection{Proof of Theorem \ref{thm_equalthr2users}}

\begin{figure}
  \centering
  \includegraphics[width=0.40\textwidth]{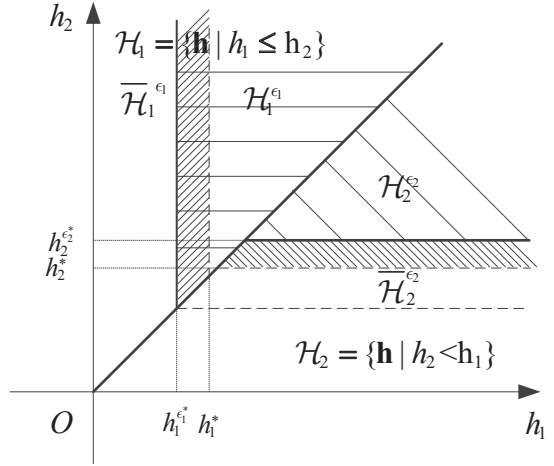}\\
  \caption{Proof of Theorem \ref{thm_equalthr2users}.}\label{fig_thm4}
\end{figure}

\begin{proof}
We prove this theorem by contradiction. Suppose that there is an optimal separation $(\epsilon^*_1, \epsilon^*_2)$ which has $h_1^{\epsilon^*_1}\neq h_2^{\epsilon^*_2}$, without loss of generality, assuming that $h_1^{\epsilon^*_1}< h_2^{\epsilon^*_2}$. Let $\gamma_1^*(\mathbf{h})$, $\gamma_2^*(\mathbf{h})$ be the corresponding optimal power allocation in subspaces $\mathcal{H}_1$, $\mathcal{H}_2$ respectively,
\begin{align} \nonumber
&\gamma^*(\mathbf{h})=\left\{\begin{array}{ll}
\gamma_1^*(\mathbf{h}),& \mathbf{h}\in\mathcal{H}_1 \\
\gamma_2^*(\mathbf{h}),& \mathbf{h}\in\mathcal{H}_2
\end{array}
\right.
\end{align}
Then, as discussed previously, in order to satisfy the service outage constraint, $R(h_1\gamma_1^*(\mathbf{h}))\geq r_0$,  $R(h_2\gamma_2^*(\mathbf{h}))\geq r_0$  in the subspaces $\mathcal{H}_{1}^{\epsilon_1}$, $\mathcal{H}_{2}^{\epsilon_2}$ respectively, as shown in Fig. \ref{fig_thm4}. Since the PDFs of $h_1$ and $h_2$ are continuous, we can find $h_1^*$ and $h_2^*$ such that $h_1^{\epsilon^*_1}\leq h_1^*\leq h_2^*\leq h_2^{\epsilon^*_2}$ and $\int_{h_1^{\epsilon_1^*}}^{h_1^*}\int_{h_1}^{\infty}f_{\mathbf{h}}(\mathbf{h})\rm{d}h_2\rm{d}h_1=\int_{h_2^*}^{h_2^{\epsilon^*_2}}\int_{h_2}^{\infty}f_{\mathbf{h}}(\mathbf{h})\rm{d}h_1\rm{d}h_2$.
If $h_1$ and $h_2$ are independent, then $\int_{h_1^{\epsilon_1^*}}^{h_1^*}f_{h_1}(h_1)(1-F_{h_2}(h_1))\rm{d}h_1=\int_{h_2^*}^{h_2^{\epsilon^*_2}}f_{h_2}(h_2)(1-F_{h_1}(h_2))\rm{d}h_2$, where $F_{h_i}(x)=\int_{0}^{x}f_{h_i}(h_i)\rm{d}h_i, i=1,2$.

Let $\Delta h_1=h_1^*-h_1^{\epsilon_1^*}$, and divide $[h_1^{\epsilon_1^*},h_1^*]$ into $n$ bins,  $[h_1^{\epsilon_1^*}+\frac{i\Delta h_1}{n},h_1^{\epsilon_1^*}+\frac{(i+1)\Delta h_1}{n}]$, $i=0, 1, \cdots, n-1$. Correspondingly, we can divide $[h_2^*,h_2^{\epsilon^*_2}]$ into $n$ bins, $[h_2^i,h_2^{i+1}], i=0,1,\cdots,n-1$, with $h_2^0=h_2^*$, $h_2^{n-1}=h_2^{\epsilon^*_2}$, such that
$\int_{h_1^{\epsilon_1^*}+\frac{i\Delta h_1}{n}}^{h_1^{\epsilon_1^*}+\frac{(i+1)\Delta h_1}{n}}f_{h_1}(h_1)(1-F_{h_2}(h_1))\rm{d}h_1=\int_{h_2^i}^{h_2^{i+1}}f_{h_2}(h_2)(1-F_{h_1}(h_2))\rm{d}h_2$.

By the first mean value theorem for integration, we can always find
$h_1^{bi}\in [h_1^{\epsilon_1^*}+\frac{i \Delta h_1}{n},h_1^{\epsilon_1^*}+\frac{(i+1) \Delta h_1}{n}], h_2^{bi}\in[h_2^i, h_2^{i+1}]$ such that
$ \int_{h_1^{\epsilon_1^*}+\frac{i\Delta  h_1}{n}}^{h_1^{\epsilon_1^*}+\frac{(i+1) \Delta h_1}{n}}f_{h_1}(h_1)(1-F_{h_2}(h_1))\rm{d}h_1=f_{h_1}(h_1^{bi})(1-F_{h_2}(h_1^{bi}))\frac{\Delta h_1}{n}$, and $\int_{h_2^i}^{h_2^{i+1}}f_{h_2}(h_2)(1-F_{h_1}(h_2))\rm{d}h_2 = f_{h_2}(h_2^{bi})(1-F_{h_1}(h_2^{bi}))(h_2^{i+1}-h_2^i)$,
$i=0,1,\cdots,n-1.$ This is possible since the PDFs of $h_1$ and $h_2$ are continuous.

Then we show that we can improve the average capacity of this broadcast channel by interchanging the power allocations in the corresponding bins, i.e., interchange the power allocations between $[h_1^{\epsilon_1^*}+\frac{i\Delta h_1}{n},h_1^{\epsilon_1^*}+\frac{(i+1)\Delta h_1}{n}]$ and $[h_2^i,h_2^{i+1}]$.

Since $\gamma_1^*((h_1^{bi},\cdot))>\gamma_2^*((\cdot,h_2^{bi})), i=0, 1, \cdots, n-1$, this is due to that $(h_1^{bi},\cdot)$ is in the service region while $(\cdot,h_2^{bi})$ is in the outage region. By a similar proof of Lemma 2 in \cite{zhang2014opawdc}, we can prove that
\small
\begin{align}
&\sum_{i=0}^{n-1}\tfrac{1}{2}\log\Big(1+\tfrac{h_1^{bi}\gamma_1^*((h_1^{bi},\cdot))}{\sigma^2}\Big)f_{h_1}(h_1^{bi})(1-F_{h_2}(h_1^{bi})\frac{\Delta h_1}{n}+ \nonumber\\
&\sum_{i=0}^{n-1}\tfrac{1}{2}\log\Big(1+\tfrac{h_2^{bi}
\gamma_2^*((\cdot,h_2^{bi}))}{\sigma^2}\Big)f_{h_2}(h_2^{bi})(1-F_{h_1}(h_2^{bi}))(h_2^{i+1}-h_2^{i})\nonumber\\
<&\sum_{i=0}^{n-1}\tfrac{1}{2}\log\Big(1+\tfrac{h_1^{bi}\gamma_2^*((\cdot,h_2^{bi}))}{\sigma^2}\Big)f_{h_1}(h_1^{bi})(1-F_{h_2}(h_1^{bi}))\frac{\Delta h_1}{n}+ \nonumber\\
&\sum_{i=0}^{n-1}\tfrac{1}{2}\log\Big(1+\tfrac{h_2^{bi}
\gamma_1^*((h_1^{bi},\cdot))}{\sigma^2}\Big)f_{h_2}(h_2^{bi})(1-F_{h_1}(h_2^{bi}))(h_2^{i+1}-h_2^{i}) \nonumber
\end{align}
\normalsize
By letting $n\rightarrow \infty$, we show that the achieved average capacity is larger by interchanging the power allocation in those selected two regions.

Besides, by interchanging the power allocations in corresponding bins, the average power constraint and the service outage constraint are satisfied. Therefore, the new power allocation policy would achieve larger average system capacity while satisfying all the constraints. This violates the assumption that $\gamma^*(\mathbf{h})$ is optimal. Thus, the optimal power allocation policy must make $h_1^{\epsilon_1^*}=h_2^{\epsilon_2^*}$. This concludes the proof.
\end{proof}

%

\bibliographystyle{IEEEtran}
\bibliography{seroutbdc}

\end{document}